# SUCCESS CRITERIA FOR IMPLEMENTING TECHNOLOGY IN SPECIAL EDUCATION: A CASE STUDY

Mohammad Fakrudeen[1,3], Mahdi H. Miraz[1,2] and Peter Excell[2]

[1]Department of Computer Science & Software Engineering, University of Hail, KSA
`(m.miraz||m.fakrudeen|)@uoh.edu.sa`
[2]Department of Computing, Glyndwr University, Wrexham, UK
`(m.miraz||p.excell)@glyndwr.ac.uk`
[3]Department of Computing and Technology, Anglia Ruskin University, UK

*ABSTRACT*

*The Kingdom of Saudi Arabia (KSA) has made a large investment in deploying technology to develop the infrastructure and resources for special education. The aims of the present research were to find out the rate of return of these investments in terms of success and, based on the findings, to propose a framework for success criteria. To achieve these aims, a mixed methodology-based research was conducted. Our study found that the use of technology in special education could not reach the desired level of implementation. We found that various success criteria such as professional experience and technology skills of special educators, administrative support, assistive hardware issues and assistive software issues, pedagogical issues, and teaching style are the key influencing factors of the implementation process.*

## 1. INTRODUCTION

Recent developments in assistive technology and ICT have heightened the need for these to be used for special education. In recent years, there has been an increasing interest in special education by the Government of the Kingdom of Saudi Arabia (KSA). In the Kingdom, more than 27000 special students (including approximately 35% female) were registered for special education in the schools in the year 2010-2011 with a teacher (special educator) to student ratio of 2:9. The percentage of special students is approximately 0.053% of the total number of students enrolled in a year [1].

The Saudi Ministry of Education (MOE) reports that the Saudi government inaugurated more than 38 audio programmes in 18 centres around the KSA for deaf students between 2006 and 2010 and more than 362 programmes for intellectually disabled and blind students at intermediate and high school level. Arabic-enabled software was also provided for teaching computers for visually impaired students by the education ministry [2].

This research is a part of iPLEASE (The Intelligent distributed Prototype by Learning, Empowering and Authoring in Special Education) which is a research project that explores ways of implementing technology in inclusive schools [3]. The focus of this paper is to study the current information and communication technology (ICT) and special education system in the KSA and reveal its issues.

This research was conducted in association with the MOE, KSA and University of Hail. The term ICT used in the whole paper indicates all the resources associated to computers.

## 2. BACKGROUND

The most prevailing themes in special education technology are assistive technology, implementation issues, instructional design, instructional strategies, outcomes of technology, professional development, reading and technology, and technology integration [4]. The schedule of Saudi schools is based on the American system [5]. In the past, parents of the students with disabilities were responsible for providing any assistance to their children [6], but at present, The MOE in KSA is responsible for providing and developing curricula, and establishing training programmes to in-service teachers for all students, including those with disabilities [7].

In 1962, the MOE established the Department of Special Learning to improve learning and rehabilitation services in three main categories of students with disabilities: those with blindness, deafness, and the intellectually disabled [8].

Legislation of Disability (LD) was passed in 1987 that guarantees individuals with disability rights equal to those of other people in society. LD requires that public agencies must provide rehabilitation services and training programmes that support independent living [9].

The disability code was also passed by the Saudi government in 2000 to assist eligible people in areas including welfare, habilitation, health, education, training and rehabilitation, employment, complementary services, and other areas [9].

The Regulations of Special Education Programs and Institutes (RSEPI) were introduced in 2001. Under the RSEPI, all children with disabilities are entitled to a free and appropriate education and individual education programmes [10]. Thus, RSEPI supports the quality of the special education services in KSA.

To summarize, the KSA has revised and passed laws to support the equal rights of individuals with disabilities in obtaining free and appropriate education. The KSA made a very large investment in ICT in special education in terms of infrastructure and resources. The Saudi Government has also made a great effort to enhance the country's educational system over the past decade by introducing new education programmes, research and development initiatives, and building numerous schools and universities [11]. The Saudi Gazette reports that the KSA is ranked as the largest educational investor in the Gulf region, with educational development projects worth $3.26 billion expecting completion in 2012 [12]. A major portion of this investment is for deploying ICT in special education as well as developing rich infrastructure and resources for it.

The question is, to what degree was this investment in the current special education system equipping its students adequately to survive in the 21st Hi-tech century? This brings us to the central objective of the present study. To what extent is the ICT technology used in special education in schools and what are the success criteria for implementing ICT in special education?

## 3. RESEARCH DESIGN

The research follows a multi-methodological approach [13] by combining qualitative [14]methods (by observing users, interviews, focus group discussions), and quantitative [15] methods (by using questionnaires, statistical analysis and experiments) together.

### 3.1. Participants

A qualitative approach was used to reveal the issues of information and communication technology in the special education system in the KSA. The multiple case-study approach was adopted to increase the reliability of data, with the help of a team of researchers and tutors to reduce bias [16][17]. In this study, the MOE chose 4 public intermediate schools for boys having special education classes in the Hail region in the KSA. From these four institutions, 50 special students, 10 special educators and 4 special education administrators have participated.

### 3.2. Research Instrument

#### 3.2.1. Questionnaire for Special Educators

The questionnaire presented to the special educators was in English. In addition, the translated version (Arabic) was also supplied if demanded. It focused on the central research variables and information about background variables at teacher and school level. It focused primarily on four particular types of ICT competencies.

1. To know the personal and professional information of special educators
2. To know the ICT skills of special educators
3. To quantify ICT usage in special education
4. To know ICT infrastructure used in special education

#### 3.2.2. One to One Interview with Special Students

For obvious reasons, the researchers were not able to communicate directly with the special students on their own. Special educators were required to assist the researchers to interact with the students. Several small tests such as simple arithmetic, language proficiency test, computer usage (opening and reading word documents using screen reader), reading test using Braille were conducted to check the students' skills and level of understanding of different subjects.

#### 3.2.3. One to One Interview with Special Education Officers

Initially, an overview of the study and a brief introduction regarding the interview were delivered to the special education officers. The formal interview process was then led by the researchers. To standardize the interview, all the participants were asked the same questions following the same sequence. Both open-ended as well as yes/no response questions were used. The participants were provided with enough information regarding the types of question (open ended or yes/no) before they were asked to answer them. Each question asked was related to the initial objectives of the research to determine factors that influence the use of technology during instructional time.

### 3.3. Scope and Limitations

All the four boys' schools included in this study are situated in an urban area of the Hail region in KSA. The most significant limitation of this study was the population included in the study. The time frame for interviewing and surveying the participants posed another limitation to this study. Given additional time, the study could include a larger, more diverse population of participants, thus creating a more complete view of the situation being studied.

## 4. RESULTS AND DISCUSSION

The study found that most of the special education institutes as well as public schools lack a multidisciplinary team, IQ tests, adaptive behaviour scales, and academic scales that appropriate to cultural standards of KSA [18]. The study conducted by Hanafi [19] examined the viability of related services for students with hearing disabilities and Quraini [20] for cognitive disabilities in public schools. This study indicated that health and medical services were more available for these students; however, social workers and rehabilitation service were not available. Our study found that ICT was not implemented properly in inclusive schools.

Although the MOE has invested heavily in the infrastructure in special education and is ready to invest more, lack of knowledge about the latest technology used in special education, lack of ICT curriculum, lack of research in special education and lack of involvement of public agencies in special education were the main factors which hindered the implementation. We have summarized our findings according to the criteria which were found to be the principal hurdles to the implementation of ICT in special education.

### 4.1. Professional Experiences of Special Educators

The study shows that almost all sample special educators were well experienced in this field. Approximately 40% special educators have around eleven years of experience and another 40% of them have nine to ten years of experience in their current role(see Figure1). Nearly 50% of special educators did not undergo any training for special education since they had their initial job induction at the time of joining. Only 50% of special educators had had the opportunity to attend any special education training in this field. If any of them face any technical problems such as using sign language, they have to consult either the experienced teachers or trained special educators. They also surf the websites to get access to that knowledge domain.

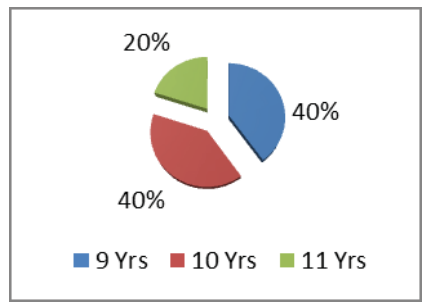

Figure 8 Special Education Experience of sampled teachers

Our study also found that although 90% of special educators are of the opinion that it would be desirable to integrate ICT in the classroom, no significant correlation was found between teachers' comfort levels in using ICT and the level of teaching experience. The research reveals that confidence in using technology among special educators relies on their pre-existing ICT profiles.

### 4.2. ICT Profile of Special Educators

All the special educators included in our survey reported to be familiar with ICT to some extent. Almost all of them are using the Internet, but mostly for checking emails. Of the study population, only 63% were aware about search engines such as Google, Yahoo and so on. 88% of the special educators had knowledge about general purpose software such as MS Office. None of the special educators knew about programming or games software used in special education. Figure 2 shows the ICT usage profile of sample special educators.

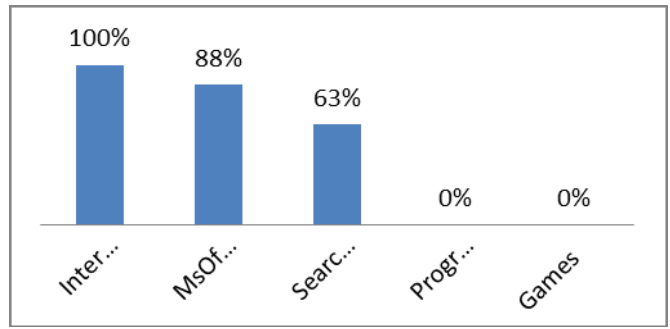

Figure 2 ICT Profile of special educators

None of the sample special educators undergoes any ICT training either by the MOE or by themselves. It can be concluded that although they have fair knowledge of ICT on their own, efforts should be made by the administrators to strengthen their ICT Skills which will facilitate integration of ICT in the classroom.

**4.3. Administrative Support**

The study also found that administrative support plays a catalyst role in integrating ICT in the classrooms. The school principals and special education officers are the actors in the catalysis process. The school principals play an important role in demanding resources in the form of hardware, software and so forth to the schools from the ministry. Then it is imperative for the special education officers to implement the ICT in the special education. MOE reveals that the administrative staff-student ratio for the academic year of 2010-2011 was 1:81 [1].

A huge factor in technology integration lies in the understanding of the relationship between administrators and their style of leadership [21]. The study found that special education officers were not in a position to implement effective ICT in their schools because their demands for implementation were not met by MOE. This clearly indicates the lack of a technology transition plan and lack of coordination between the MOE and the special education officers.

We summarize the following criteria for preparation for the implementation of ICT from an administrative point of view:

1. Periodic meetings to be conducted between special education officers and the MOE to discuss regarding required support and infrastructure in implementing ICT in their school.
2. Regular meetings to be scheduled between special education officers and special educators to analyse implementation of ICT based on pedagogy.
3. Special education officers have to select special educators for training in ICT and peer coaching.
4. Flexible schedule to be provided to special educators in order to increase their ICT skills
5. Technology transition plan to be developed based on a number of special students and special educators

The study also reveals that special education officers are not updated about the latest advances in the field of assistive hardware, assistive software and other ICT resources used in special education. Hence they are unable to recommend suitable ICT tools available in the market for the schools.

## 4.4. Assistive Hardware Issue

A computer-oriented educational technology programme includes several components, such as hardware, software, training, support, and maintenance [22]. The study found that infrastructure varies among the schools depending on the number as well as the need of the special students. In some schools, a separate laboratory was assigned for special students, but in others only common laboratories were used for inclusive education.

The study reveals that available ICT infrastructure was not properly utilized to teach technical subjects such as science, mathematics and also linguistic subjects such as Arabic and English. In brief, special educators were not properly using the ICT to design, plan and deliver their lessons to the special students in the focus group.

Therefore, we conclude that the following criteria affect implementation of ICT effectively from an assistive hardware point of view:

1. Non-availability of technical person to assist special educators in case of failure in hardware
2. Failure in strategy for reducing effective computer-student ratio
3. Non-applicability of ICT in specific subjects such as mathematics and, science in computer labs
4. Difficulties in sharing hardware resources among general and special students in inclusive classrooms
5. Lack of availability of updated hardware resources such as switches and adaptive keyboard.

## 4.5. Assistive Software Issue

The study also investigated assistive software used in the classroom. Our research reveals that in some schools, ICT was used at the basic level to practice numbers and alphabets on the keyboard. Most of the schools were not using game software for edutainment purpose in special education, whereas this is a powerful tool for special needs education.

Special educators feel difficulty in using screen readers such as JAWS and Super nova in interaction with special students. While all the schools have the medium of instruction in Arabic, most of the screen readers were application-specific and were able to read only in English. In some schools, Arabic screen readers were used, but they are not common in all the schools. The MOE has issued a CD in sign language in both English and Arabic, but it has been found that none of the special educators either use it to educate themselves or to educate their special students.

To conclude, the following are the criteria which hinder implementation of effective ICT in the classroom with respect to assistive software:

1. Lack of matching software according to curriculum
2. Lack of resources which affects the usage of software
3. Lack of software available in Arabic version
4. Lack of knowledge regarding available freeware for the special educators to download and use in the classroom.
5. As Saudi education system is region-oriented, it was difficult to find suitable region-specific educational software available in the market.

## 4.6. Pedagogical Factors

A skill test was conducted as part of the study to find pedagogical factors affecting implementation of ICT in the classroom. The skill test was conducted in subjects such as mathematics and English for hearing impaired students at intermediate level. It was found that hearing impaired students aged between 12 and 14 were able to do simple arithmetic but they were unable to perform complex arithmetic.

Hearing impaired students were also tested for English language skill. It was found that they were unable to form a complete sentence, which indicates lack of grammar skill.

The researchers also reviewed the teaching methodology adopted by special educators for cognitive disabilities. It has been found that the ministry has not framed either a curriculum or issued any course textbooks for elementary, intermediate and secondary levels for cognitive students. MOE has set only objectives for each level. Special educators were adopting methodologies from other countries to teach cognition-impaired students.

Visually impaired students were also tested for mathematical and computer skills. The research revealed that all subjects were taught through Braille for the visually impaired. Students were able to solve all the problems supplied by the researchers.

To summarize, the following criteria are required to implement ICT effectively from the pedagogical point of view:

1. It is necessary to re-evaluate current curriculum and course textbooks in order to boost confidence to the special students and enable them to pursue higher education and stand independently in the future

2. The ICT-based curriculum has to be developed. For instance, curriculum could be split using the key areas which can accommodate computer, internet and digital media using screen reader.

3. Special students ought to be encouraged to use ICT by providing assessments and e-books in order to use ICT in their daily life.

But monitoring the implementation of these criteria is also required since, as argued by Goodison [23], the definition of a national ICT curriculum on its own does not guarantee any instructional use of ICT. An interesting issue in the context of this discussion was the balance between the extrinsic and intrinsic forces that drive the integrated use of ICT by teachers. Imposed policy decisions were often less responsive to teacher perspectives and often neglected workplace constraints [24].

## 4.7. Teaching Style

Curriculum reform, based on ICT, is therefore unlikely to succeed unless we understand teachers' personal perspectives and teaching style. The Saudi education system is based on inclusive education where general and special students study in the same school. After completion of school studies, disabled students can join vocational training and employment skills that support independent living [2]. However, while research conducted by Al-Faiz [25] suggested that most of the teachers have positive attitudes toward inclusive education, our research found a completely different scenario. The sample schools which were surveyed reported that the same subjects were taught to general students and visually impaired students together in one classroom and hearing-impaired students in another classroom. Special educators are of the opinion that it would be difficult to combine the hearing impaired students with the general students. This clearly indicates that the purpose of inclusive education was not fulfilled as expected.

The research reveals that teaching style of special educators is mostly instruction oriented. They are more willing to take the challenge of using technology only when they know that there is someone else in the vicinity to help, guide or correct them should a mishap occur. The research also reveals that usage of ICT in the classroom by special educators also depends on the confidence level of special students to use the technology. It was found that at the elementary level, special students use less technology than general students because of their disability. But at the intermediate and secondary level, special students were more enthusiastic in using technology. We also found that special students at the intermediate and secondary level were using social networking sites and they made groups of their own choice. It clearly demonstrates that ICT has become a part of their daily life.

## 5. CONCLUSIONS

As KSA continues its dramatic period of improvement, usage of ICT in special education will increase rapidly. This paper provides insight views regarding the current status of ICT in special education in the KSA. In addition, it emphasizes the challenges and limitations that were encountered in use of ICT in the inclusive classrooms for teaching the students with disability. The success criteria, suggested in this paper, could also be applied to other Middle East countries which share the similar cultural and social as well as economic structure.

Furthermore, this research will serve as a foundation for future studies and the suggestions set forth in this paper might contribute towards implementing ICT effectively in special education. Returning to the questions posed at the beginning of this study, it is now possible to state that, given the need for the curriculum to be framed based on ICT, then existing ICT skills of special educators have to be improved. The teaching styles have to be more technology oriented. Before implementing ICT in schools, transition plans are supposed to be developed and periodic meetings conducted between various groups who influence the implementation of ICT in the classroom. The latest technology in assistive hardware and assistive software ought to be made available in the schools by the MOE. It was noteworthy that special students were using social networking sites and initiating groups of their own. This demonstrates that they are becoming 'digital natives': it is a common phenomenon that students may be more digitally able than their teachers and hence mentoring of the teachers to accommodate this and not suppress the ability and enthusiasm of the students is very important for the future.

Finally, individuals with disabilities need to be treated equally in the society. They also require more attention at school. Use of ICT and assistive technology is a big step towards achieving this. Since the government is very concerned about the special students, the situation will soon change dramatically. By following the above success criteria, it is to be hoped that the special students can pursue higher education to stand independently and succeed in whatever they see their future to be.